\documentclass[10pt,a4paper]{article}
\usepackage{opex3}
\usepackage{color}
\usepackage{threeparttable}

\usepackage{amsmath,amsthm,amsfonts,amssymb,cite}
\usepackage{graphicx,array,MnSymbol}
\usepackage[usenames,dvipsnames]{xcolor}

\newcommand{\lp}{\left(}  
\newcommand{\rp}{\right)} 
\newcommand{\lb}{\left[}   
\newcommand{\rb}{\right]}  

\newcommand{\mr}{\mathrm} 
\newcommand{\mb}{\mathbf} 
\newcommand {\cl}{\mathcal} 
\newcommand{\vectsym}{\boldsymbol}  
\DeclareMathOperator{\Real}{Re\,}     

\DeclareMathOperator{\upd}{\,d} 

\newcommand{\brho}{\vectsym{\rho}} 
\newcommand{\bd}{\mb{d}} 

\newcommand{\abs}[1]{\left | #1 \right |}   

\begin{document}

\title{Interferometric superlocalization of two incoherent optical point sources}

\author{Ranjith Nair$^{1,*}$ and Mankei Tsang$^{1, 2}$}

\address{$^1$Department of Electrical and Computer Engineering, \\
National University of Singapore, 4 Engineering Drive 3, \\Singapore 117583\\
$^2$Department of Physics, \\
National University of Singapore, 2 Science Drive 3, Singapore 117551}

\email{$^*$elernair@nus.edu.sg} 



\begin{abstract}
A novel interferometric method -- SLIVER (Super Localization by Image inVERsion interferometry) -- is proposed for  estimating  the separation of two incoherent point sources with a mean squared error that does not deteriorate as the sources are brought closer. The essential component of the interferometer is an image inversion device that inverts the field in the transverse plane about the optical axis, assumed to pass through the centroid of the sources.  The performance of the device is analyzed using the Cram\'er-Rao bound applied to the statistics of spatially-unresolved photon counting using photon number-resolving and on-off detectors. The analysis is supported by Monte-Carlo simulations of the maximum likelihood estimator for the source separation, demonstrating the superlocalization effect for separations well below that set by the Rayleigh criterion. Simulations indicating the robustness of SLIVER to mismatch between the optical axis and the centroid are also presented. The results are valid for any imaging system with a circularly symmetric point-spread function.
\end{abstract}

\ocis{(110.2970) Image detection systems; (350.5730) Resolution; (110.0180) Microscopy; (030.5290)   Photon statistics; (180.2520) Fluorescence Microscopy; (180.3170) Interference Microscopy; (180.1790) Confocal Microscopy; (180.5810) Scanning Microscopy.} 


\section{Introduction}
Rayleigh's criterion for resolution of two incoherent point sources, which asserts that a minimum separation between the sources equal to the diffraction-limited spot size is necessary for them to be resolvable, has perhaps been the most influential resolution criterion in the history of optics despite its heuristic character \cite{Ray1879xxxi,BW99Principles}. 
Based as it is on the eye as a detection instrument, Rayleigh's criterion neglects the possibility of better resolution using light of greater intensity or using a longer observation period. Pioneering studies on resolution limits as a function both of the source separation and the mean number of observed photons were made by Bettens \emph{et al.} in \cite{BDD+99} and more recently by Ram \emph{et al.} in \cite{RWO06} (See also \cite{VDD+02} in the context of electron microscopy). In \cite{BDD+99}, it was shown that for a given mean photon number, ideal continuum image-plane photon counting can locate the centroid, i.e., the midpoint, of the two radiating sources with a finite  precision (depending on the size of the diffraction-limited spot in the image plane and on the number of photons detected) no matter how small their separation is. Based on the Cram\'er-Rao (CR) lower bound of estimation theory \cite{VanTreesI}, it was also shown in \cite{BDD+99,RWO06} that any (unbiased) estimate of the separation between the sources based on image-plane photon counting must suffer a divergent mean squared error for a given mean photon number as the separation tends to zero. This phenomenon was dubbed  \emph{Rayleigh's curse} in \cite{TNL15} as it suggests a fundamental limitation in resolving incoherent point sources even when the role of the average detected photon number is taken into account. In the past few decades, several  far-field super-resolution techniques that circumvent the Rayleigh limit have revolutionized microscopy (see \cite{WS15} for a review), but these rely on placing restrictions on the emissions from the sources and do not fundamentally challenge the criterion (or Rayleigh's curse) for independently and incoherently radiating sources. Other developments, e.g.,\cite{CRW+13}, use clever techniques to approach the performance of ideal continuum image-plane photon counting using imperfect detectors.

In \cite{TNL15}, following preliminary work in \cite{Tsa15},  the problem of resolving two incoherent point sources was approached anew from the perspective of quantum estimation theory using the quantum Cram\'er-Rao (QCR) bound \cite{Hel76,Hol11,Par09}. This bound provides a fundamental limit to the accuracy of estimating the source separation optimized over all possible measurement techniques allowed by quantum mechanics.  Under a weak-source assumption, it was found in \cite{TNL15} that the QCR lower bound on the minimum mean squared error (MSE) of estimating the separation of two point sources is \emph{independent} of that separation. Further, a linear optics-based measurement -- spatial-mode demultiplexing (SPADE) -- was proposed and shown in principle to approach the quantum bound for all values of the separation. These results are in stark contrast to the performance of image-plane photon counting mentioned above, as the divergent behavior of the minimum MSE with decreasing separation -- Rayleigh's curse -- is conspicuously absent. 

In this paper, we revisit the resolution problem from the point of view of the semiclassical theory of photodetection \cite{MW95,Sha09,Goo85Statistical}. For sources emitting thermal radiation, for linear propagation of  light through the imaging system, and for linear-optics processing followed by photon counting, the semiclassical theory of photodetection may be used to obtain the same results as a fully quantum analysis would give  \cite{MW95,Sha09}. This is because the thermal state has a positive $P$-representation, and its counting statistics may be obtained by statistical averaging over those of coherent states via Mandel's rule \cite{MW95}. The SLIVER scheme of this paper as well as the schemes of \cite{TNL15} satisfy the above necessary conditions for the applicability of the semiclassical theory. We emphasize that, for the state of radiation and measurements considered here and in \cite{TNL15}, ``semiclassical'' does not imply ``approximate'' -- the fully quantum and semiclassical treatments are in quantitative agreement. 

Besides its greater familiarity to the optics community, a semiclassical analysis of the resolution problem offers several advantages. Firstly, the analysis can be carried out for arbitrary source strengths. Indeed, the work in \cite{BDD+99,RWO06} assumes the sources are weak enough so that the counting statistics are nearly Poissonian. Similarly, the quantum analysis in \cite{TNL15} assumed, for mathematical tractability, that the sources are weak enough so that the state of light could be assumed to be confined to the zero and one-photon subspaces.  Thus, a semiclassical treatment will incorporate the  effects of multi-photon events neglected in \cite{BDD+99,RWO06} and in the quantum analysis of \cite{TNL15}. Such an analysis would also reveal whether the superresolution effect predicted in \cite{TNL15} for weak sources persists as the source strength is increased. A second advantage of the semiclassical approach is that intuitions from semiclassical optics can be harnessed to both understand the reasons for superresolution as well as to  design new measurement schemes which can also be analyzed semiclassically provided that they only involve linear optics and photodetection, possibly along with the use of other classical sources of light.

Accordingly, our contributions in this paper are three-fold:-
\begin{enumerate}
\item  We set up the problem of resolving two incoherently radiating point sources in the  framework of semiclassical photodetection theory. The sources can be of arbitrary strength and imaging systems with two-dimensional circularly-symmetric point-spread functions are studied. 
The thermal source model is applicable to a wide range of physical scenarios ranging from optical astronomy to fluorescence microscopy.

\item We propose  a new interferometric scheme for estimating source separation that we call SLIVER (Super Localization by Image inVERsion interferometry) and that yields finite resolution for arbitrarily small values of the source separation and for arbitrary source strengths.  Devices employing image inversion have been proposed, studied and demonstrated previously in the microscopy literature for various related applications \cite{SG06,WH07,WSH09,Wei+11,WBK+11,WEB+13,WBK+14,WBK+15}, but the fundamental limits on the capability of such devices for improving lateral resolution seem not to have been explored before. Through an analysis of the photodetection statistics of SLIVER in the semiclassical framework, we show both using the Cram\'er-Rao bound and through explicit simulation that our method manifestly alleviates Rayleigh's curse. 

\item We offer an explanation for the superlocalization effect of SLIVER at small values of separation in the language of estimation theory applied to the photodetection statistics for thermal light. 
\end{enumerate}

\section{Source and system model}
Consider the depiction in Fig.~1 of the focused image of two incoherent point sources in the image plane of an imaging system coordinatized by $\brho = (x,y)$. We assume that the imaging system is spatially invariant and coordinates have been rescaled so that the image is of unit magnification \cite{Goo05Fourier}. We also assume that the (possibly complex-valued) normalized amplitude point-spread function (PSF) $\psi(\brho)$ satisfying
\begin{align}
\int \upd \brho\, \abs{\psi(\brho)}^2 =1
\end{align}  is circularly symmetric, so that $\psi(\brho)$ depends only on the magnitude $\abs{\brho}$ 
for all $\brho$.  It is straightforward to make $\psi(\brho)$ spatially invariant and circularly symmetric in an imaging system using two lenses \cite{Goo05Fourier}. The prototypical example of such a PSF is of course the Airy disk pattern resulting from a circular pupil. The images of the sources 1 and 2 in the image plane are assumed to be centered at $\mp \frac{\bd}{2}$ respectively, giving rise in the image plane to a combined field  with complex amplitude
\begin{align} \label{ipf}
E(\brho) = A_1\psi\lp\brho+\frac{\bd}{2}\rp + A_2\psi\lp\brho-\frac{\bd}{2}\rp.
\end{align} 
in normalized units of $\sqrt{\mbox{photons}\cdot\mbox{m}^{-2}}$. We will assume throughout this paper, except in Section 5.3, that  the centroid of the sources, i.e., their midpoint,  has already been located, perhaps by image-plane photon counting \cite{RWO06}. It is taken to lie at the origin of the image plane. In Sec.~5.3, we  consider the effect of a small error in centroid localization. Our initial results suggest that our measurement scheme is robust to such perturbations. 

The pair of dimensionless complex numbers $A = (A_1,A_2) \in \mathbb{C}^2$ are the source amplitudes. The thermal and mutually incoherent nature of the sources dictates that the source amplitudes are circular-complex Gaussian random variables having the first and second moments \cite{Goo85Statistical}:-
\begin{align} 
&\mathbb{E}[A_{\mu}] =  0 \label{S1}\\
&\mathbb{E}[A_{\mu}A_{\nu}] = 0 \label{S2}\\
&\mathbb{E}[A_{\mu}^*A_{\mu}] = \epsilon_\mu \label{S3} \\
&\mathbb{E}[A_{1}^*A_2] = 0 \label{S4}
\end{align}
for $\mu,\nu \in \{1,2\}$ ranging over the two sources. According to the above relations, the real and imaginary parts of the $\{A_{\mu}\}$ comprise four statistically independent zero-mean Gaussians of variance $\epsilon_1/2$ (for the components of $A_1$) or $\epsilon_2/2$ (for the components of $A_2$). Eqs.~(\ref{S1})-(\ref{S4}) are as appropriate for completely incoherent sources with respective strengths (i.e., mean photon number) $\epsilon_1$ and $\epsilon_2$, which can assume any values. 

\begin{figure}[h]
\centering\includegraphics[trim=80mm 80mm 80mm 42mm, clip=true,scale=0.5]{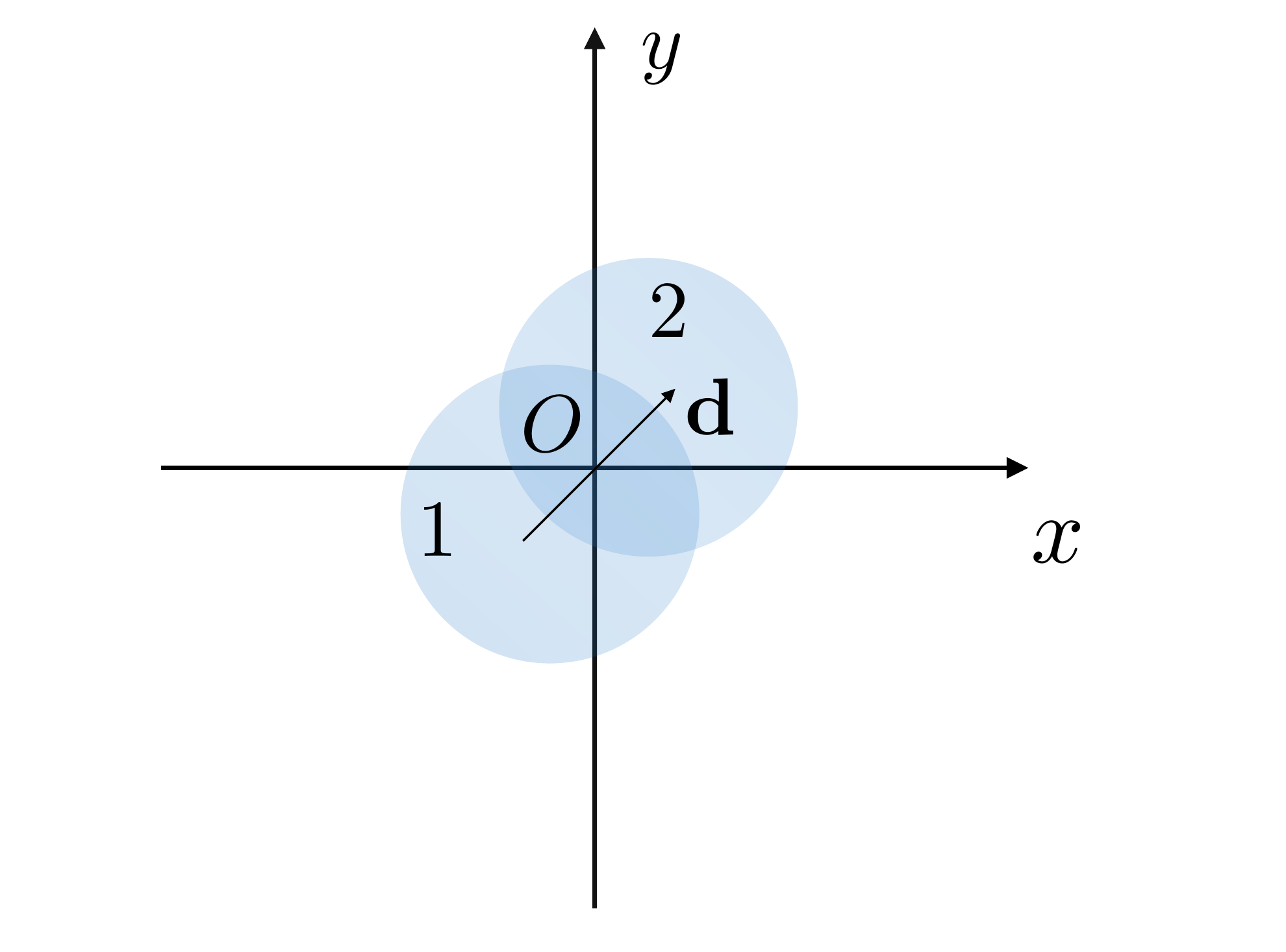}
\caption{A rendering of the focused image of two point sources $1$ and $2$ located at the tail and tip respectively of the vector $\bd$. The blue spots indicate the approximate extent of the point-spread function (PSF) centered at each source. The centroid of the sources is assumed to coincide with the intersection $O$ of the optical axis with the image plane. The focus of this paper is the regime in which $d = \abs{\bd}$ is smaller than the width of the PSF, in which conventional image-plane photon counting performs poorly in estimating $d$. }
\end{figure}

Given that the centroid has been located, the parameter of interest is  the separation  $d =\abs{\bd}$ between the two sources. The angle between the line joining the sources and the $x$-axis is another unknown parameter, but the  measurement method proposed here works regardless of the value of that angle, which need not be known beforehand in practice. We will thus focus just on the separation $d$ and apply the tools of single-parameter estimation theory in our analysis.

\section{The SLIVER measurement and its statistics}
The SLIVER measurement, illustrated schematically in Fig.~2 and described in detail below, essentially involves the separation of the image-plane field  $E(\brho)$ of Eq.~\eqref{ipf} into its \emph{symmetric} and \emph{antisymmetric} parts with respect to  inversion in the image plane about the optical axis. By definition, the symmetric and antisymmetric parts of $E(\brho)$ are given respectively by
\begin{align} \label{Esa}
\begin{split}
E_s(\brho) &:= \frac{E(\brho)+E(-\brho)}{2} = \frac{S}{2}\lb \psi\lp\brho + \frac{\bd}{2}\rp + \psi\lp\brho - \frac{\bd}{2}\rp\rb  \\
E_a(\brho) &:= \frac{E(\brho)-E(-\brho)}{2} = \frac{D}{2}\lb\psi\lp\brho + \frac{\bd}{2}\rp- \psi\lp\brho - \frac{\bd}{2}\rp\rb.
\end{split}
\end{align}
In writing the right-hand sides of Eqs.~\eqref{Esa}, we have used the fact that $\psi(\brho)$ is circularly symmetric and have defined the sum and difference
\begin{align}
S &= A_1 + A_2 \\
D &= A_1 - A_2
\end{align}
of the amplitudes of the two sources. The relations
\begin{align} \label{SD}
\begin{split}
&\mathbb{E}[S] = \mathbb{E}[D] = 0 \\
&\mathbb{E}[S^2] =\mathbb{E}[D^2] = \mathbb{E}[SD] = 0 \\
&\mathbb{E}[S^{*} S] = \mathbb{E}[D^{*} D] = \epsilon_1 + \epsilon_2 \\
&\mathbb{E}[S^*D] = \epsilon_1 - \epsilon_2
\end{split}
\end{align}
hold for these new random variables, which are thus also identically distributed circular-complex Gaussians and statistically independent of each other when $\epsilon_1 = \epsilon_2$. 

The field components of Eq.~\eqref{Esa} may be obtained in spatially separated planes using an \emph{image inversion interferometer} (III) that works by splitting the input field using a 50-50 beamsplitter, spatially inverting the field from one output about the optical axis and recombining the two beams at a second 50-50 beamsplitter.  The two beams output from the second beamsplitter then have the field patterns of Eq.~\eqref{Esa}. The entire setup is thus essentially a balanced Mach-Zehnder interferometer with an inversion device in one arm (Fig.~2). The critical element in such setups is the inversion device -- a device that implements the transformation $E_\mr{out}(\brho) = E_\mr{in}(-\brho)$. Inversion devices have been proposed and demonstrated for various applications in the fluorescence microscopy community and a considerable literature on image inversion microscopy exists \cite{SG06,WH07,WSH09,Wei+11,WBK+11,WEB+13,WBK+14,WBK+15}. However, to the best of our knowledge,  a fundamental statistical analysis of the kind we make here of the capabilities of an III for lateral resolution has not been done (See the discussion in Section 6). The spatial inversion itself can be accomplished in a variety of ways, e.g., using a $4f$-arrangement with lenses \cite{WH07}, with a 3-D setup using plane mirrors \cite{WSH09}, or using two concave mirrors \cite{Wei+11}. We will be concerned here only with the outputs \eqref{Esa} of such a device rather than its detailed implementation, which may vary according to the application. 

 Suppose now that the output beams from the III are directed to different photodetectors. 
Conditioned on the values of $S$ and $D$, the intensity patterns $I_s(\brho_s)$  and $I_a(\brho_a)$ of the fields on their respective detector planes are
\begin{align} 
I_s(\brho_s) &= \frac{\abs{S}^2}{4}\lb \abs{\psi\lp \brho_s + \frac{\bd}{2}\rp}^2 + \abs{\psi\lp \brho_s - \frac{\bd}{2}\rp}^2 +2I_{\mr{int}}(\brho_s, \bd) \rb, \\
I_a(\brho_a) &= \frac{\abs{D }^2}{4}\lb \abs{\psi\lp \brho_a + \frac{\bd}{2}\rp}^2 + \abs{\psi\lp \brho_a - \frac{\bd}{2}\rp}^2  -2I_{\mr{int}}(\brho_a, \bd) \rb, 
\end{align} 
where
\begin{align}
I_{\mr{int}}(\brho, \bd) =\Real  \psi^*\lp \brho + \frac{\bd}{2}\rp\psi\lp \brho - \frac{\bd}{2}\rp
\end{align}
is an interference term.
Conditioned on the values of $S$ and $D$, semiclassical detection theory  dictates that the photocounts on the two detector planes are independent  inhomogeneous Poisson processes with the rate functions
$I_s(\brho_s)$ and $I_a(\brho_a)$  respectively \cite{MW95,Goo85Statistical,Sha09}. It follows that the spatially-unresolved integrated photocounts $N_s$ and $N_a$ at each detector are Poisson random variables with the means

\begin{align} \label{sameans}
\mathbb{E}[N_s | S] &=  \frac{\abs{S}^2}{2} \int \upd \brho_s I_s(\brho_s)  = \frac{\abs{S}^2}{2}\lb 1+ \delta(d) \rb, \\
 \mathbb{E}[N_a | D] &=  \frac{\abs{D}^2}{2} \int \upd \brho_a I_a(\brho_a) = \frac{\abs{D}^2}{2}\lb 1- \delta(d) \rb,
\end{align}
where
\begin{align}
\int \upd \brho I_{\mr{int}}(\brho, \bd) &= \Real \int \upd\brho\,\psi^*(\brho)\psi(\brho-\bd)  \label{delta}\\
&=\Real \iint\limits_{-\infty}^{\infty} \upd x \upd y \,\psi^*(x,y)\psi(x-d,y) \label{E2}\\
&\equiv \delta(d).
\end{align}
The circular symmetry of $\psi(\brho)$ has been used to get Eq.~\eqref{E2}, and we have retained only the separation $d$ as argument. 
According to Eqs.~\eqref{SD}, both $\abs{S}^2$ and $\abs{D}^2$ are  exponentially distributed with mean $(\epsilon_1+ \epsilon_2)$, so the photocounts $N_s$ and $N_a$ integrated over the photodetector surfaces are distributed according to Bose-Einstein distributions \cite{MW95,Goo85Statistical,Sha09} with means
\begin{align}
\bar{N}_s &= \frac{\epsilon_1+\epsilon_2}{2} \lb 1 +\delta(d)\rb = \epsilon_{\mr{ave}} \lb 1 +\delta(d)\rb, \\
 \bar{N}_a &= \frac{\epsilon_1 + \epsilon_2}{2} \lb 1 -\delta(d)\rb= \epsilon_{\mr{ave}} \lb 1 -\delta(d)\rb, \label{Na}
\end{align}
for $\epsilon_{\mr{ave}} = (\epsilon_1 + \epsilon_2)/2$. If $\epsilon_1 = \epsilon_2$ , the photocounts are also statistically independent because $S$ and $D$ are independent in that case.

The SLIVER method consists of optimal statistical processing of the sequence of observed photocount measurements of $N_s$ and (or) $N_a$ over $M$ independent shots using the same sample to obtain a good estimate $\hat{d}$ of the separation $d$ -- see Fig.~2 and Secs.~4 and 5. As we show below, the photocount from the antisymmetric component of the input field is much more informative in this regard. 
\begin{figure}[!htbp]
\centering\includegraphics[trim=0mm 0mm 0mm 0mm, clip=true, width=0.9\columnwidth]{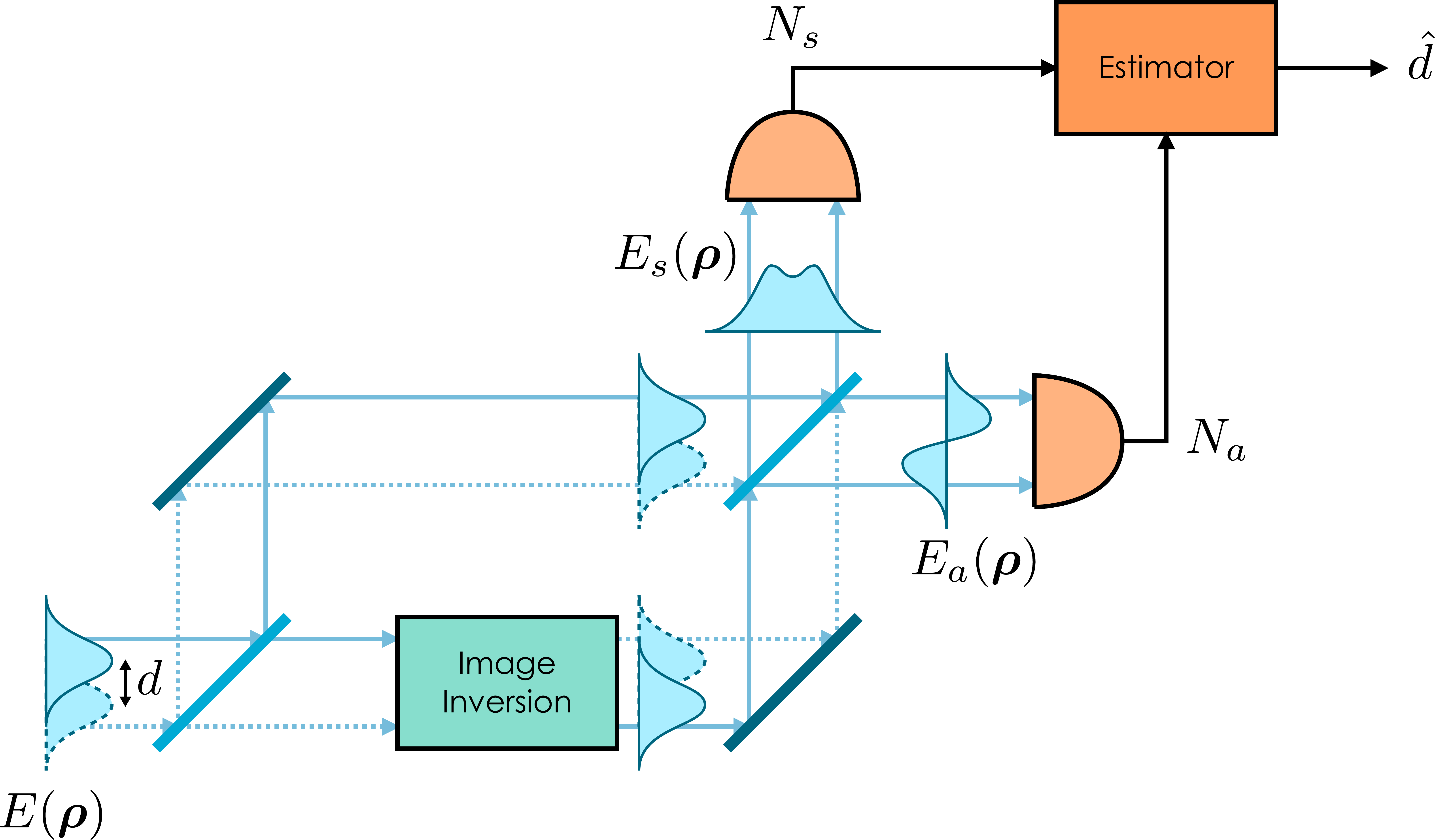}
\caption{The SLIVER method for separation estimation:-- The input field $E(\brho)$ is separated into its symmetric ($E_s(\brho)$) and antisymmetric ($E_a(\brho)$) components with respect to inversion about the centroid of the sources, which is also the optical axis. This separation is accomplished by an image inversion interferometer as shown, where the image inversion box can be implemented by any of a variety of methods (see Sec.~3).  The components impinge upon separate bucket detectors that collect spatially-unresolved photocounts $N_s$ and $N_a$. The photocount record over a series of $M$ detection windows of $N_s$ and (or) $N_a$ is processed to yield an estimate $\hat{d}$ of the separation $d$. }
\end{figure}

\section{Cram\'{e}r-Rao bounds for the separation $d$}
\subsection{Number-resolved photon counting}
Given the Bose-Einstein distribution 
\begin{align} \label{Ps}
P_s(n) &:=\mbox{Pr}[N_s = n] = \frac{1}{\bar{N}_s +1}\lp\frac{\bar{N}_s }{\bar{N}_s+1} \rp^n,
\end{align}
of number-resolved photon counting of the inversion-symmetric field, we can calculate the fundamental Cram\'{e}r-Rao (CR) lower bound of estimation theory \cite{VanTreesI} on the mean squared error (MSE) $\mathbb{E}[\hat{d}(N_s) -d]^2$ of any unbiased estimator $\hat{d}(N_s)$ of $d$ that is a function of the photocount $N_s$. An unbiased estimator is one that satisfies $\mathbb{E}[\hat{d}(N_s)] = d$ for all $d$ in some interval of interest. The statistical expectations here are taken over the distribution of $N_s$. The CR bound is itself the reciprocal of the Fisher information $\cl{J}_d^{\mr{(sym-pc)}}$ on the parameter $d$ defined as
\begin{align}
\cl{J}_d^{\mr{(sym-pc)} }= \mathbb{E}_{N_s}\lb \frac{\partial \log P_s(n)}{\partial d} \rb^2.
\end{align}
The superscript indicates that the  Fisher information pertains to photon counting of the inversion-symmetric field. This quantity turns out to equal 
\begin{align} 
\cl{J}_d^{\mr{(sym-pc)} }& = \sum_{n=0}^{\infty} P_s(n) \lb \frac{\partial \log P_s(n)}{\partial d} \rb^2\\
 & = \lb  \frac{\partial  \bar{N}_s}{\partial d} \rb^2 \sum_{n=0}^{\infty} P_s(n) \lb \frac{\partial \log P_s(n)}{\partial \bar{N}_s} \rb^2 \\
&=    \lb  \frac{\partial  \bar{N}_s}{\partial d} \rb^2 \, \frac{1}{\bar{N}_s (\bar{N}_s +1)} \label{septerms}\\
&= \frac{\epsilon_\mr{ave}\gamma^2(d)}{1 +\delta(d)} \left\{1 + \epsilon_\mr{ave}\lb 1 + \delta(d)\rb\right\}^{-1}. \label{Jspc}
\end{align}
Here  we have defined 
\begin{align} \label{gamma}
\gamma(d) = \delta'(d) =- \Real \iint\limits_{-\infty}^\infty \upd x \upd y\, \psi^*(x,y) \frac{\partial \psi(x-d,y)}{\partial x}.
\end{align}
In a similar fashion, we find the Fisher information of photon counting of the inversion-antisymmetric field to be
\begin{align} \label{Japc}
\cl{J}_d^{(\mr{asym-pc})} &= \frac{\epsilon_\mr{ave}\gamma^2(d)}{1 -\delta(d)} \left\{1 + \epsilon_\mr{ave}\lb 1 - \delta(d)\rb\right\}^{-1}.
\end{align}
The CR lower bounds for these individual measurements then respectively read
\begin{align}
\mathbb{E}[\hat{d}(N_s) -d]^2 &\geq \frac{1}{\cl{J}_d^{\mr{(sym-pc)} }} = \frac{1+\delta(d)}{\epsilon_\mr{ave}\gamma^2(d)} + \lb \frac{1+\delta(d)}{\gamma(d)}\rb^2 , \\
\mathbb{E}[\hat{d}(N_a) -d]^2 &\geq \frac{1}{\cl{J}_d^{\mr{(asym-pc)} }} =  \frac{1-\delta(d)}{\epsilon_\mr{ave}\gamma^2(d)} + \lb \frac{1-\delta(d)}{\gamma(d)}\rb^2. 
\end{align}
For a fixed $d$, both bounds contain a term that displays the familiar shot-noise scaling with respect to the combined source strength $\epsilon_\mr{ave}$. Of greater interest here, however, is the behavior of the bounds as the separation reduces to zero for fixed source strengths.
Since $\gamma(d) \rightarrow 0$ and $\delta(d) \rightarrow 1$ as $d \rightarrow 0$, we have
\begin{align}
\lim_{d \rightarrow 0}   \frac{1+\delta(d)}{\gamma^2(d)} = \infty,
\end{align}
so that any unbiased estimator based on bulk direct detection of the inversion-symmetric field suffers a divergent mean squared error for separations $d \rightarrow 0$. Thus, Rayleigh's curse that plagues spatially-resolved image-plane photon counting  remains in effect here \cite{BDD+99,RWO06}.  However, note that
\begin{align}
\lim_{d \rightarrow 0} \frac{ 1 -\delta(d)}{ \gamma^2(d)} = -\frac{\delta'(0)}{2\gamma(0) \gamma'(0)} = - \frac{1}{2\gamma'(0)} \equiv \frac{1}{2(\Delta k)^2}.
\end{align}
Here we have used
\begin{align}
\gamma'(d)& =- \Real \iint\limits_{-\infty}^\infty \upd x \upd y\,\frac{\partial \psi^*(x+d,y)}{\partial x} \frac{\partial \psi(x,y)}{\partial x}, \\
\gamma'(0) & = - \iint\limits_{-\infty}^\infty \upd x \upd y\,\abs{\frac{\partial \psi(x,y)}{\partial x}}^2 \\
& \equiv - (\Delta k)^2, \label{Deltak2}
\end{align}
where $ (\Delta k)^2$ is the squared spectral width of the PSF. Remarkably therefore, direct detection of the \emph{inversion-antisymmetric field does not suffer Rayleigh's curse} and a limiting MSE
\begin{align} \label{MSEmin}
\frac{1}{M\cl{J}_0^{\mr{(asym-pc)} }} =  \frac{1}{2M\epsilon_\mr{ave}(\Delta k)^2} \equiv \frac{1}{N(\Delta k)^2}
\end{align}
is potentially achievable for small $d$, where $M$ is the number of independent measurements, and
\begin{align} \label{N}
N= 2M\epsilon_\mr{ave} 
\end{align}
is the the mean number of photons emitted over the $M$ shots. The limiting Fisher information value $2M \epsilon_\mr{ave}(\Delta k)^2$ is similar to the quantum limit obtained in \cite{TNL15} and achieved by the SPADE and binary SPADE schemes therein. An advantage of SLIVER over those schemes  is that there is no need to couple the image-plane field into a waveguide(s) with mode profile(s) tailored to the PSF. 

If $\epsilon_1=\epsilon_2$, $N_s$ and $N_a$ are statistically independent so that the two Fisher information terms of eqs.~\eqref{Jspc} and \eqref{Japc} may be added to give the total Fisher information. The reciprocal of this quantity is then the CR bound for any unbiased estimate $\hat{d}(N_s,N_a)$ of $d$ based on both photocounts. In the remainder of the paper, however, we will  only consider measurements in the \emph{antisymmetric} output port.

\subsection{On-off detection of the inversion-antisymmetric field}

Consider bucket direct detection at the antisymmetric port with an \emph{on-off detector}, i.e., a detector that only distinguishes between no photon and one or more photons. The no-click probability and click probability are respectively
\begin{align}
P_a(0) &= \frac{1}{1+ \epsilon_{\mr{ave}}\lb 1- \delta(d)\rb} \\
P_a(>0) &= \frac{\epsilon_{\mr{ave}}\lb 1- \delta(d)\rb}{1+ \epsilon_{\mr{ave}}\lb 1- \delta(d)\rb}.
\end{align}
The Fisher information for this binary measurement can be shown to be
\begin{align} \label{Joo}
\cl{J}^{\mr{(asym-oo)}}_d &= \frac{\epsilon_{\mr{ave}}\gamma^2(d)}{ 1 -\delta(d)} \left\{1 + \epsilon_{\mr{ave}}\lb 1 - \delta(d)\rb\right\}^{-2} 
 < \cl{J}_d^{(\mr{asym-pc})},
\end{align}
which is expected as on-off detection is a coarser measurement than photon counting. However, it is interesting that on-off detection also evades Rayleigh's curse for small $d$ and the Fisher information attains the optimal value at $d=0$. Note also that the Fisher information of Eq.~\eqref{Joo} is negligibly different from that of Eq.~\eqref{Japc} for $\epsilon_\mr{ave} \ll 1$. These behaviors are expected as multi-photon arrivals in the image-plane are rare in the regime $\epsilon_\mr{ave} \ll 1$ while multi-photon detections at the antisymmetric output port are rare in the $\epsilon_{\mr{ave}}\lb 1- \delta(d)\rb \ll 1$ regime, i.e., the regime of $d \approx 0$. Equations~\eqref{Japc} and \eqref{Joo} indicate that the two antisymmetric detection methods should exhibit superresolution for \emph{arbitrary} values of source strength beyond the $\epsilon_\mr{ave} \ll 1$ regime.

\begin{figure}[!htbp] 
\centering\includegraphics[trim=20mm 80mm 30mm 80mm, clip=true,width=0.7\textwidth]{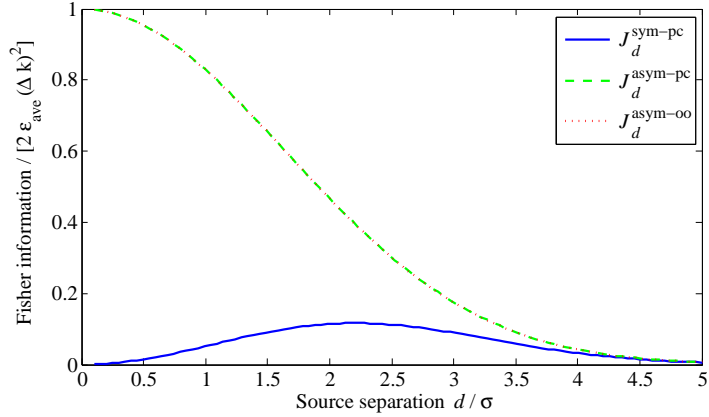}
\caption{Fisher information obtainable using the SLIVER method: -  The Fisher information for photon counting $J_d^{\mr{sym-pc}}$ at the symmetric output port of the image inversion interferometer (blue curve), photon counting at the antisymmetric output port of the image inversion interferometer ($J_d^{\mr{asym-pc}}$ -- dashed green curve) and for on-off detection at the antisymmetric output port ($J_d^{\mr{asym-oo}}$ -- dotted red curve) are plotted against the source separation $d$. The plots are normalized with respect to the maximum value $2 \epsilon_\mr{ave} (\Delta k)^2$ of $J_d^{\mr{asym-pc}}$ and $J_d^{\mr{asym-oo}}$, attained at $d=0$. The superlocalization effect, consistent with the quantum bound of \cite{TNL15}, is evident for $d \sim 0$. The circular Gaussian PSF of Eq.~\eqref{Gaussianpsf} has been assumed and the plots are independent of the value of the half-width $\sigma$. The average source strength $\epsilon_\mr{ave}= 10^{-3}$ photons.} 
\end{figure}
 We illustrate the above results for the circular Gaussian PSF
\begin{align} \label{Gaussianpsf}
\psi_G(\brho) = \frac{1}{(2\pi\sigma^2)^{1/2}}\exp\lp-\frac{\abs{\brho}^2}{4\sigma^2}\rp.
\end{align}
The PSF-dependent quantities appearing in the CR bound then become:-
\begin{align}
\delta_G(d) &= \exp\lp-\frac{d^2}{8\sigma^2}\rp, \\
\gamma_G(d) &= -\frac{d}{4\sigma^2}\,\exp\lp-\frac{d^2}{8\sigma^2}\rp, \\
(\Delta k)_G^2 &= \frac{1}{4\sigma^2}.
\end{align}
The Fisher information quantities of Eqs.~\eqref{Jspc}, \eqref{Japc}, and \eqref{Joo} pertaining to photon counting at the symmetric and antisymmetric ports, and on-off detection at the antisymmetric port of the interferometer are plotted as a function of the separation $d$ in Figures 3-4. The source strength is $\epsilon_\mr{ave}=10^{-3}$ photons and $0.5$ photons in Figures 3 and 4 respectively. The plots are normalized to the maximum value of $J_d^{\mr{asym-pc}}$  attained at $d=0$. We see that the information obtainable from the antisymmetric detection methods is maximum at $d=0$ and decreases thereafter. The information from detection at the symmetric port is zero for $d=0$ and reaches a peak around $d=2-2.5 \sigma$ and decreases thereafter.

\begin{figure}[htbp]
\centering\includegraphics[trim=20mm 80mm 30mm 80mm, clip=true, width=0.7\textwidth]{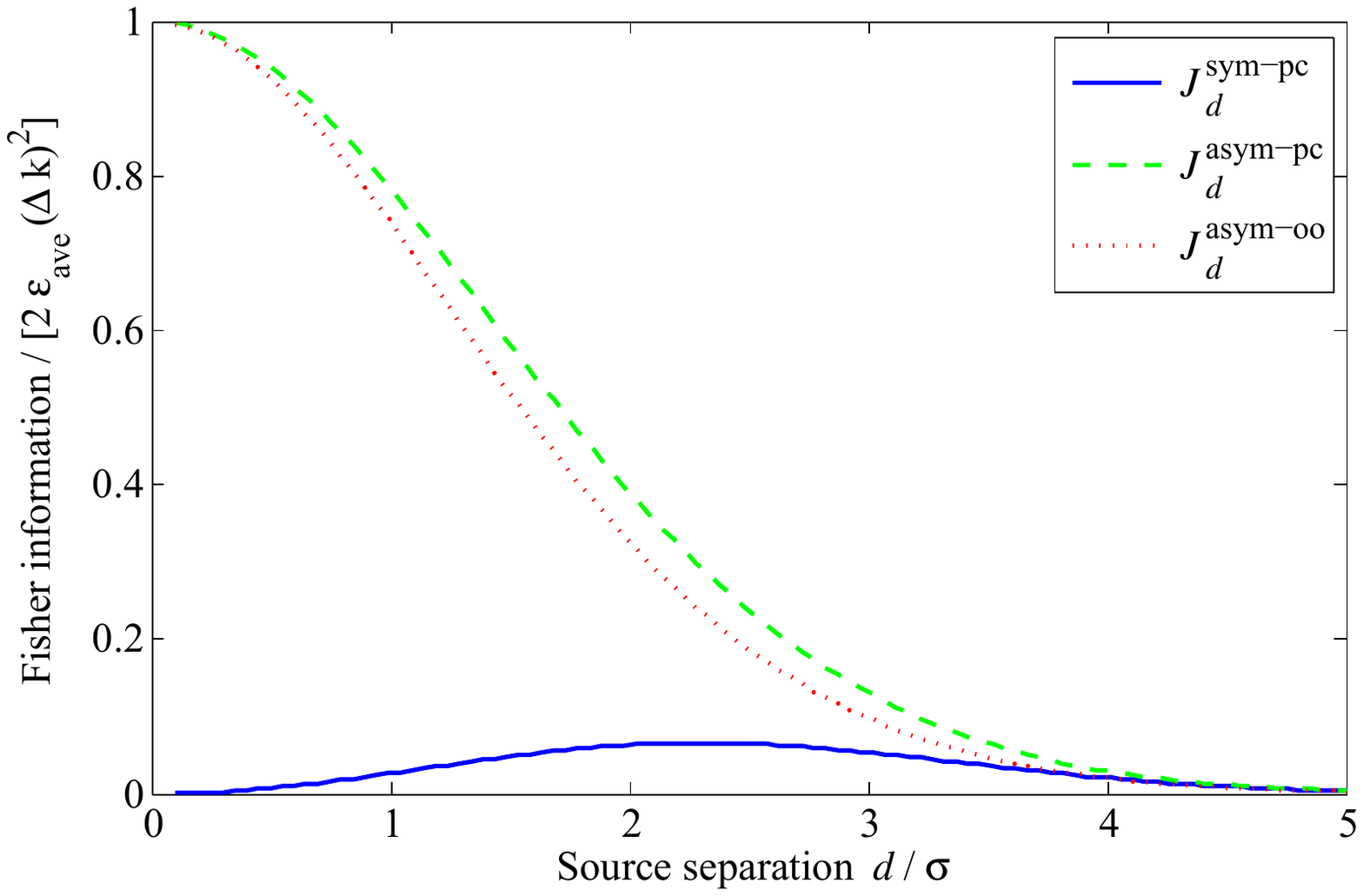}
\caption{Fisher information obtainable using the SLIVER method: -  The Fisher information for photon counting $J_d^{\mr{sym-pc}}$ at the symmetric output port of the image inversion interferometer (blue curve), photon counting at the antisymmetric output port of the image inversion interferometer ($J_d^{\mr{asym-pc}}$ -- dashed green curve) and for on-off detection at the antisymmetric output port ($J_d^{\mr{asym-oo}}$ -- dotted red curve) are plotted against the source separation $d$. The plots are normalized with respect to the maximum value $2 \epsilon_\mr{ave} (\Delta k)^2$ of $J_d^{\mr{asym-pc}}$ and $J_d^{\mr{asym-oo}}$, attained at $d=0$. The superlocalization effect, consistent with the quantum bound of \cite{TNL15}, is evident for $d \sim 0$. The circular Gaussian PSF of Eq.~\eqref{Gaussianpsf} has been assumed and the plots are independent of the value of the half-width $\sigma$. The average source strength $\epsilon_\mr{ave}= 0.5$ photons, so that superresolution persists outside the regime $\epsilon_\mr{ave} \ll 1$ and is more marked for number-resolved measurements.} 
\end{figure}

\section{Maximum-Likelihood (ML) estimates of $d$: Monte-Carlo analysis}
To further support the predictions from the CR bounds, we present Monte-Carlo simulations of the MSE for on-off detection and photon counting 
at the antisymmetric output port of the interferometer. The circular Gaussian PSF of Eq.~\eqref{Gaussianpsf} is assumed and the simulation results are independent of the half-width $\sigma$.

\subsection{On-off detection in the antisymmetric output port}
Consider direct detection of the output $E_a(\brho)$ in $M$ detection windows using an on-off detector. The ensuing measurement record is a bit string $\mb{k} = (k_1, \ldots, k_M)$, with $k_m=0$ if the detector did not fire in the $m$-th detection window and $k_m=1$ if it did. The maximum likelihood (ML) estimator  \cite{VanTreesI} for $d$ is then:-
\begin{align} \label{MLestimatoroo}
\hat{d}_\mr{ML} = \left\{
	\begin{array}{ll}
		 2 \sigma \sqrt{-2 \ln \lp 1 - \frac{K}{(M-K) \epsilon_\mr{ave}}\rp}  & \mr{if}\;\; \frac{K}{M-K} < \epsilon_\mr{ave}.\\
		 2 \sigma & \mbox{otherwise.}
	\end{array}
\right.
\end{align}
Here, $K = \sum_{m=1}^M k_m$ is the total number of clicks observed in the detector during the measurement, which is a sufficient statistic for generating the estimate.
The second case above is necessary because the equation for the maximum likelihood estimate of $d$ has no solution if $K/(M-K) \geq \epsilon_\mr{ave}$, in which case the estimate is set to an arbitrary value. As $M$ increases, the probability of such ``large deviations'' goes to zero.

\begin{figure}[htbp]
\centering\includegraphics[trim=20mm 80mm 30mm 80mm, clip=true,width=\textwidth]{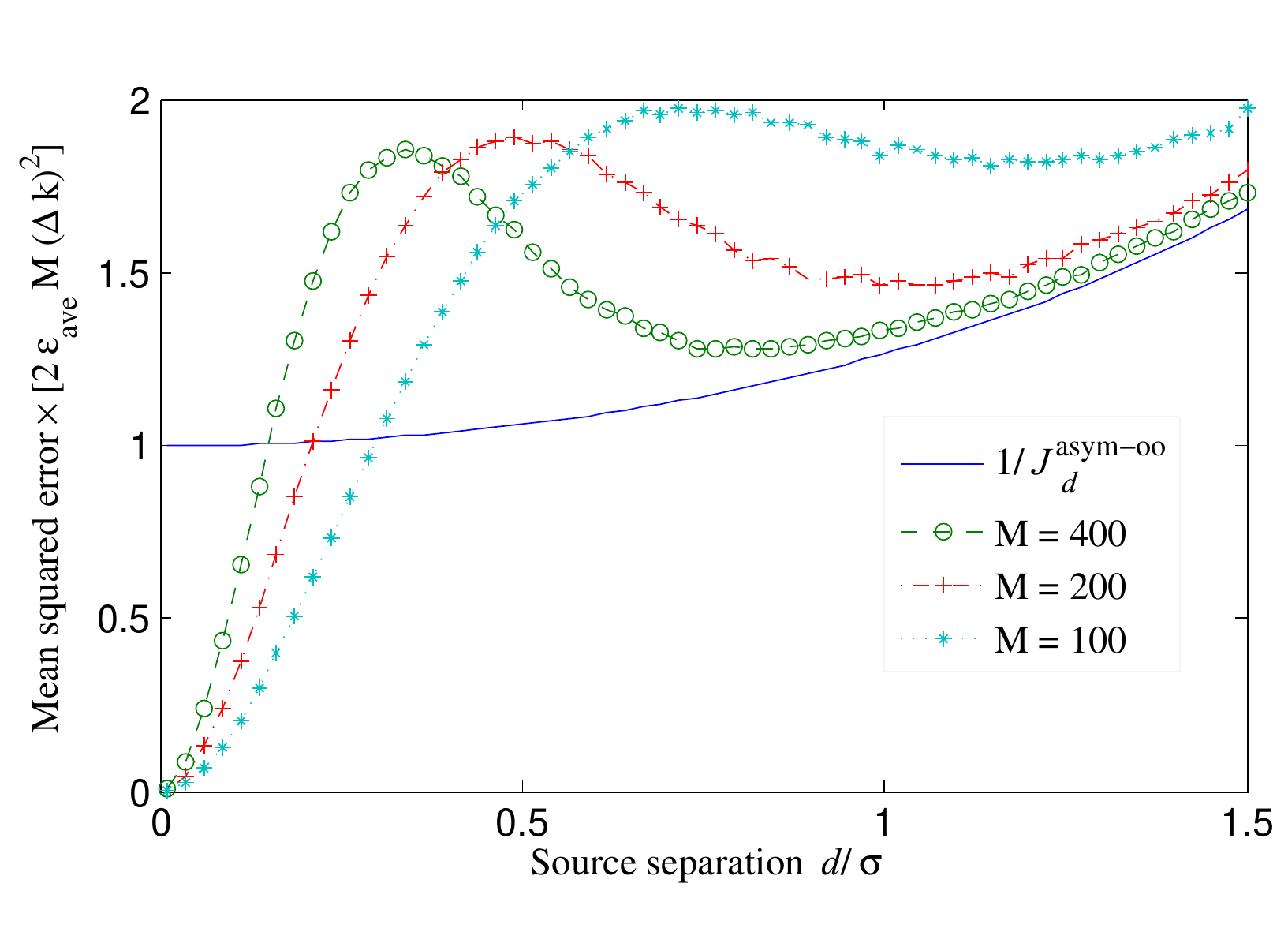}
\caption{Simulated mean squared errors for the maximum likelihood estimator $\hat{d}_\mr{ML}$ of Eq.~\eqref{MLestimatoroo} as a function of the source separation for $M = 100, 200, 400$ measurements. Each measurement consists of a binary value indicating whether or not an on-off detector in the antisymmetric output port of the interferometer registered a click.  The MSE curves are each scaled by $[2M\epsilon_\mr{ave} (\Delta k)^2]^{-1}$, the value of the Cram\'er-Rao bound for that $M$ at $d=0$. Each data point was obtained as an average of $10^5$ Monte-Carlo runs. The Cram\'er-Rao bound for on-off detection in the antisymmetric output port (solid blue curve) normalized to unity at its minimum value of $[2M\epsilon_\mr{ave} (\Delta k)^2]^{-1}$ is also shown. A circular Gaussian PSF (Eq.~\eqref{Gaussianpsf}) was assumed  and the source strength $\epsilon_\mr{ave} = 0.2$ photons. The simulated results are independent of the half-width $\sigma$.}
\end{figure}

Figure 5 shows the simulated MSE of the ML estimate for a source strength of $\epsilon_\mr{ave} = 0.2$ photons. The MSE is shown scaled relative to the minimum value $[2M\epsilon_\mr{ave} (\Delta k)^2]^{-1}$ of the CR bound for $M$-values of $100$, 200 and $400$. As the separation increases beyond the half-width, the curves for different $M$ approach the Cram\'er-Rao bound for on-off detection in the antisymmetric output port. We see that the ML estimate actually beats the CR bound for small $d \lesssim 0.3 \sigma$. This is because the ML estimate is biased for finite $M$, as may be expected from the highly nonlinear nature of the estimator of Eq.~\eqref{MLestimatoroo}. We recall here that bias and MSE are independent characteristics of an estimator. Many minimum MSE estimators are biased \cite{VanTreesI}, and techniques for reducing MSE by introducing bias are well-known in the signal processing literature \cite{KE08}. Since MSE is the relevant metric for most applications, the behavior of the MSE observed here is a welcome -- though intriguing -- feature.  For all values of the separation, the performance -- when not superior to that predicted by the CR bound --  is within a factor of 2 of it. The behavior of the simulated MSE of SLIVER with respect to the CR bound is very similar to that reported for the SPADE and binary SPADE schemes in ~\cite{TNL15}. 

For PSFs other than the circular Gaussian, the ML estimator takes a different form than that in Eq.~\eqref{MLestimatoroo}, but depends on the measurement result only through the quantity $\lp \frac{K}{M-K} \rp$.

\subsection{Photon counting in the antisymmetric output port}
For number-resolved photon counting in the antisymmetric output port, the measurement record $\mb{p} = (p_1,\ldots, p_M)$  is now a vector of non-negative integers. The total number of photons counted  -- $P = \sum_{m=1}^M p_m$ -- is a sufficient statistic and the ML estimator for $d$ is:-
\begin{align} \label{MLestimatorpc}
\hat{d}_\mr{ML} = \left\{
	\begin{array}{ll}
		 2 \sigma \sqrt{-2 \ln \lp 1 - \frac{P}{M \epsilon_\mr{ave}}\rp}  & \mr{if}\;\; \frac{P}{M} < \epsilon_\mr{ave}.\\
		 2 \sigma & \mbox{otherwise,}
	\end{array}
\right.
\end{align} 
where the last value is again an arbitrary assignment needed in the event that a solution to the equation for the ML estimate for $d$ does not exist.  

\begin{figure}[htbp]
\centering\includegraphics[trim=20mm 80mm 30mm 80mm, clip=true,scale=0.75]{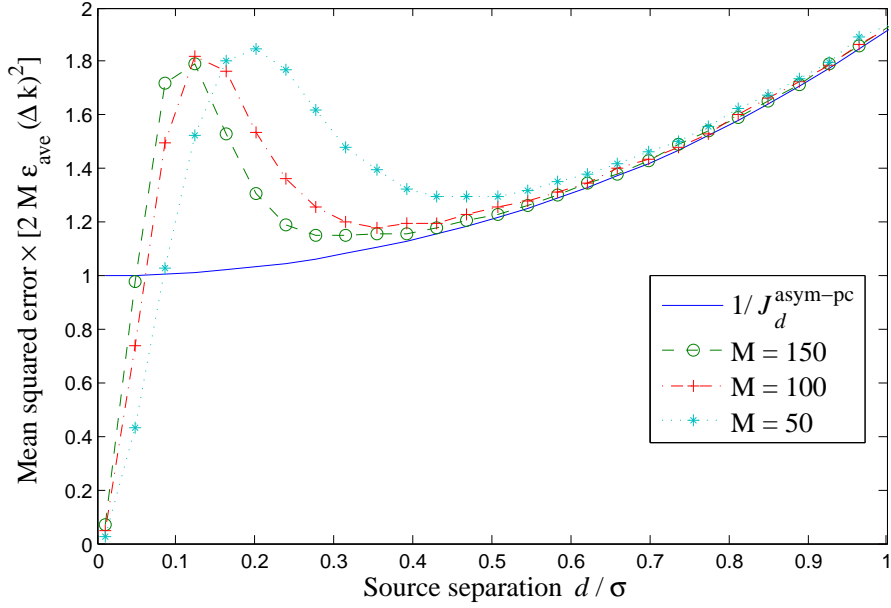}
\caption{Simulated mean squared errors for the maximum likelihood estimator $\hat{d}_\mr{ML}$ of Eq.~\eqref{MLestimatorpc} as a function of the source separation for $M = 50, 100, 150$ measurements. Each measurement counts the number of photons  in the antisymmetric port, i.e., is a measurement of $N_a$ of Sec.~2. The MSE curves are each scaled by $[2M\epsilon_\mr{ave} (\Delta k)^2]^{-1}$, the value of the Cram\'er-Rao bound for that $M$ at $d=0$. Each data point was obtained as an average of $10^5$ Monte-Carlo runs. The Cram\'er-Rao bound for photon counting in the antisymmetric output port (solid blue curve) normalized to unity at its minimum value of $[2M\epsilon_\mr{ave} (\Delta k)^2]^{-1}$ is also shown. A circular Gaussian PSF (Eq.~\eqref{Gaussianpsf}) was assumed, the source strength $\epsilon_\mr{ave} = 5$ photons, and the simulated results are independent of the half-width $\sigma$.
}
\end{figure}
Figure 6 shows the results for the MSE of the ML estimate for a source strength  $\epsilon_\mr{ave} = 5$ photons, well outside the $\epsilon_\mr{ave} \ll 1$ regime. For such a source, an on-off detector would need a very large $M$ to collect useful statistics because of detector saturation. The simulated MSE is  shown scaled relative to the minimum value $[2M\epsilon_\mr{ave} (\Delta k)^2]^{-1}$ of the CR bound for $M$-values of 50, 100 and 150. The ML estimate again beats the CR bound for small $d \lesssim 0.1 \sigma$ owing to the biasedness of the estimator of Eq.~\eqref{MLestimatorpc}. However,  the curves for each $M$ closely approach the CR bound for photon counting in the antisymmetric output port for separations $d \gtrsim 0.5\sigma$. For all values of the separation, the observed performance -- when not superior to that predicted by the CR bound --  is within a factor of 2 of it.  

For PSFs other than the circular Gaussian, the ML estimator takes a different form than that in Eq.~\eqref{MLestimatorpc}, but depends on the measurement result only through the quantity $\lp \frac{P}{M} \rp$.

\subsection{Effect of centroid mismatch: On-off detection in the antisymmetric output port}
\begin{figure}[h]
\centering\includegraphics[trim=80mm 80mm 80mm 42mm, clip=true,scale=0.5]{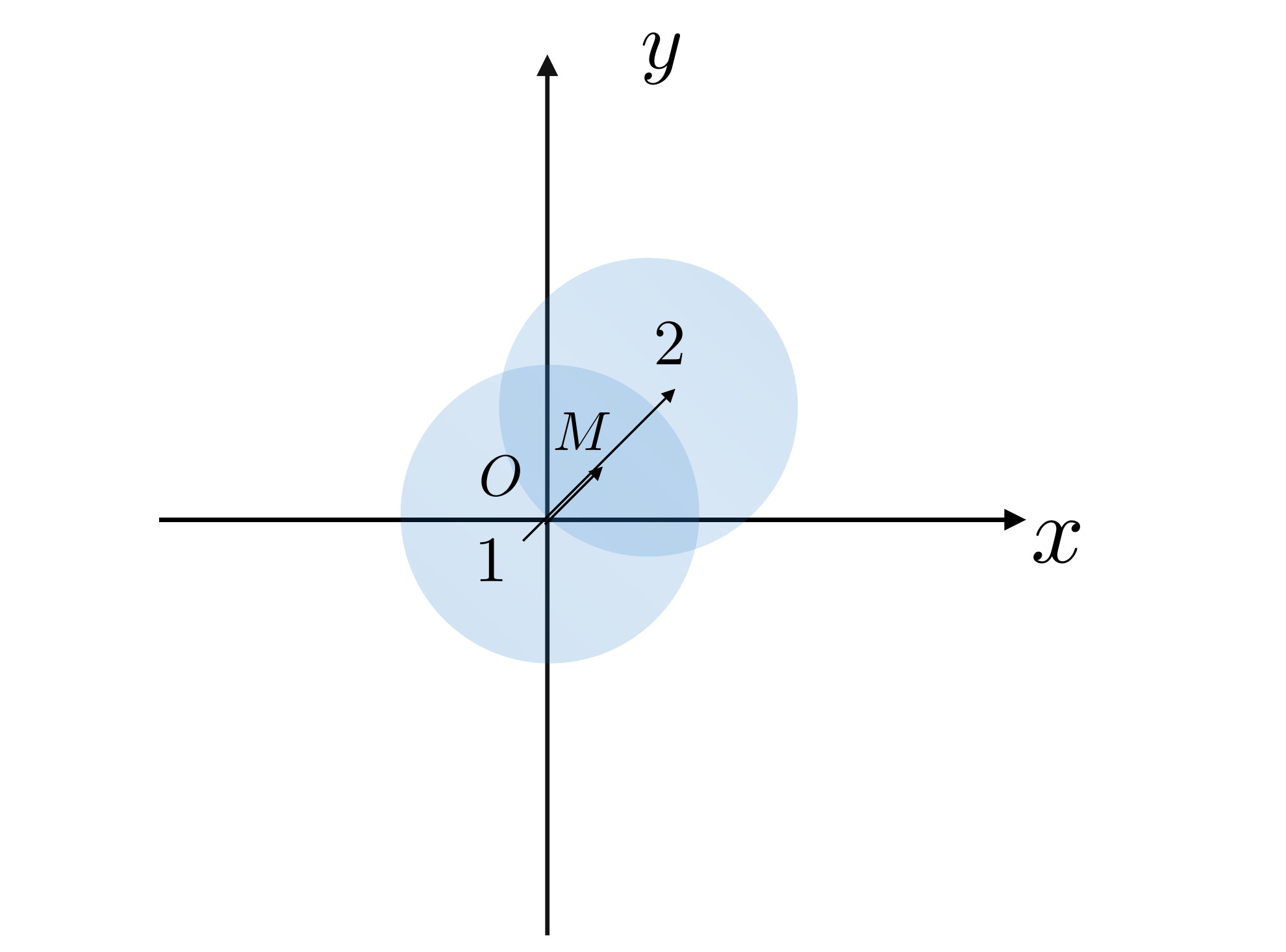}
\caption{The case of centroid mismatch considered in Section 5.3: The vector ${OM} = \bf{c}$ from the optical axis to the centroid is nonzero but parallel to the separation vector $\mathit{12}=\bf{d}$. The ordering  of the points ($\mathit{1}-O-M-\mathit{2}$) shown is illustrative --  the simulation results are valid even if $O$ does not lie between $\mathit{1}$ and $\mathit{2}$.}
\end{figure}
We have so far assumed that the centroid has been located at the optical axis, so that the vector $\bf{c}$ from the origin of the image plane to the midpoint of $\bf{d}$ is zero (Fig.~1). A detailed study of the operation and performance of SLIVER  when this assumption is removed will be given elsewhere. Here, we briefly evaluate the performance of on-off detection in the antisymmetric port for nonzero $\bf{c}$. It can be shown that the no-click and click probability are given by 
\begin{align} \label{onoffmismatch}
P_a(0) &= \lb 1 + \alpha \epsilon_1 + \beta \epsilon_2 +  \lp \alpha \beta - \gamma^2 \rp \epsilon_1 \epsilon_2\rb^{-1}, \\
P_a(>0) &= 1 - P_a(0), \nonumber
\end{align}
where
\begin{align} 
\alpha &= \frac{1-\delta\lp \abs{2\bf{c} - \bf{d}}\rp}{2}, \nonumber\\
\beta &= \frac{1-\delta\lp \abs{2\bf{c} + \bf{d}}\rp}{2}, \label{albetgam}\\
\gamma & = \frac{\delta\lp\abs{\bf{d}} \rp - \delta\lp \abs{2\bf{c}}\rp}{2}. \nonumber
\end{align}

Consider the case of $\bf{c}$  parallel to the separation vector $\bf{d}$ (Fig.~7). For the Gaussian PSF of Eq.~\eqref{Gaussianpsf}, let $\abs{\bf{c}} = \xi \sigma,$ so that $\xi$ is a normalized misalignment  factor. For  such a configuration, and for fixed $\bf{c}$ with $\xi = 0.1$, simulation results for the MSE of the estimator of Eq.~\eqref{MLestimatoroo} are shown in Fig.~8. Note that the estimator used cannot depend on the centroid position, as this is unknown in practice. We have assumed $\epsilon_1 = \epsilon_2 = \epsilon_\mr{ave} = 0.2$ photons.
\begin{figure}[htbp]
\centering\includegraphics[trim=20mm 80mm 30mm 80mm, clip=true,scale=0.7]{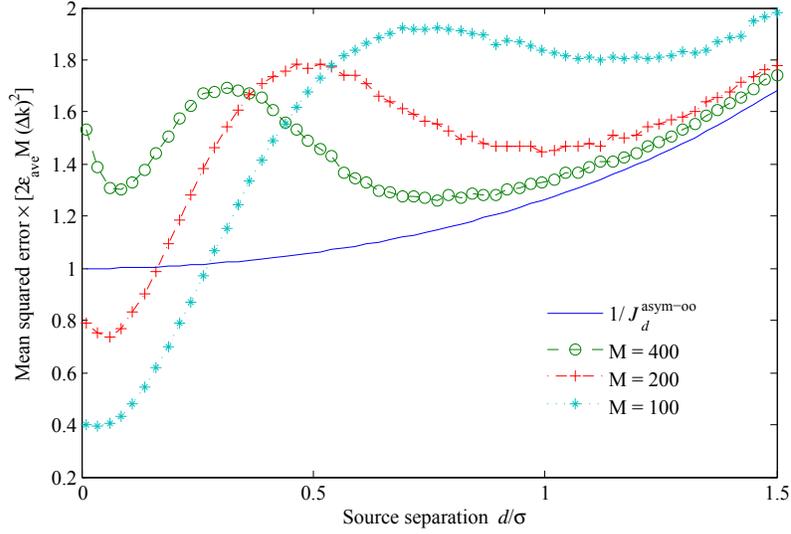}
\caption{Simulated mean squared errors for the  estimator  Eq.~\eqref{MLestimatoroo} for a misaligned SLIVER system using on-off detection in the antisymmetric port shown as a function of the source separation for $M = 100, 200, 400$ measurements. We have assumed that the centroid is a distance $\abs{\bf{c}}\equiv \xi \sigma$ away from the inversion axis, that $\bf{c}$ and $\bf{d}$ are parallel, and that $\xi=0.1$.  The MSE curves are each scaled by $[2M\epsilon_\mr{ave} (\Delta k)^2]^{-1}$, the value of the Cram\'er-Rao bound for the \emph{ideal} aligned measurement ($\xi =0$) and that $M$ at $d=0$. Each data point was obtained as an average of $10^5$ Monte-Carlo runs. The Cram\'er-Rao bound for on-off detection in the antisymmetric output port  of an aligned SLIVER system (solid blue curve) normalized to unity at its minimum value of $[2M\epsilon_\mr{ave} (\Delta k)^2]^{-1}$ is also shown. A circular Gaussian PSF (Eq.~\eqref{Gaussianpsf}) was assumed  and the source strengths $\epsilon_1= \epsilon_2 = \epsilon_\mr{ave} = 0.2$ photons. The simulated results are independent of the half-width $\sigma$.}
\end{figure}
Comparing to Fig.~5, we see that the misalignment has little effect on the MSE for separations $d \gtrsim 0.5 \sigma$. For smaller separations, the MSEs in Fig.~8 are greater than those in Fig.~5 and the MSE does not go to zero for zero separation. However, the MSEs still beat the CR bound for the \emph{aligned} SLIVER system, suggesting they  also beat the CR bound -- yet to be determined -- for the misaligned system. 

Figure 9 displays the simulated MSEs for the same source and system parameters, with $M=100$ and for the misalignment factor values $\xi = 0.1, 0.2, 0.3$ and 0.4. We see that the performance degrades gracefully with increasing misalignment, remains within an order of magnitude of the CR bound of the aligned system, and does not display the divergent behavior of image-plane photon counting \cite{BDD+99,RWO06,TNL15}.

A detailed study of SLIVER under misalignment -- including derivations of the formulas in this section, Cram\'er-Rao bounds, and performance analyses -- will be given elsewhere. Here, we simply note the important point that the MSE of a misaligned SLIVER system is still finite for arbitrarily small separations, and Rayleigh's curse has not returned.
\begin{figure}[htbp]
\centering\includegraphics[trim=20mm 80mm 30mm 80mm, clip=true,scale=0.72]{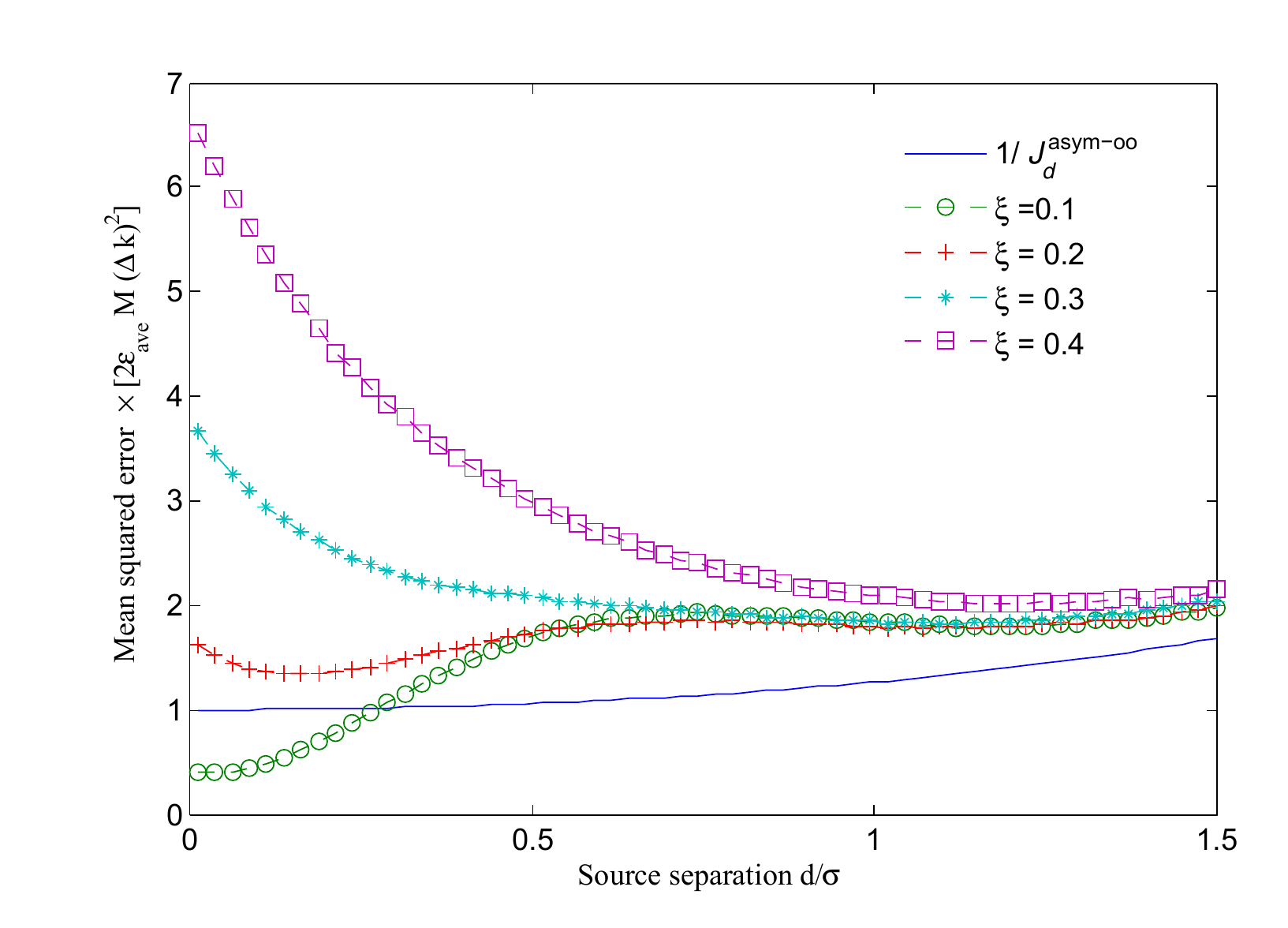}
\caption{Simulated mean squared errors for the  estimator  Eq.~\eqref{MLestimatoroo} for  misaligned SLIVER  as a function of the source separation for $M = 100$ measurements and for misalignment factors $\xi = 0.1, 0.2, 0.3$ and 0.4.  The MSE curves are each scaled by $[2M\epsilon_\mr{ave} (\Delta k)^2]^{-1}$, the value of the Cram\'er-Rao bound for the \emph{ideal} aligned measurement ($\xi =0$) at $d=0$. Each data point was obtained as an average of $10^5$ Monte-Carlo runs. The Cram\'er-Rao bound for on-off detection in the antisymmetric output port  of an aligned SLIVER system normalized to unity at its minimum value  is also shown. A circular Gaussian PSF (Eq.~\eqref{Gaussianpsf}) was assumed  and $\epsilon_1= \epsilon_2 = \epsilon_\mr{ave} = 0.2$ photons. The simulated results are independent of the half-width $\sigma$.}
\end{figure}

\section{Discussion and conclusions}
Motivated by the greater familiarity and facility of analysis afforded by the semiclassical theory of photodetection over the full quantum theory, we have set up the problem of resolving two incoherently radiating point sources in this framework. This theory is quantitatively exact for such sources and for measurements involving only linear-optics processing prior to photodetection. The  sources are assumed to be of arbitrary strength and circularly-symmetric PSF's directly related to real imaging systems are considered. Since the analysis in \cite{TNL15} was largely confined to a one-dimensional model of the point-spread function with a weak-source assumption, the model used here is of interest both theoretically and in practice (See also the remarks in Sec.~V therein). 
We have not made full semiclassical analyses of the binary SPADE and SPADE protocols of \cite{TNL15} here, as these are rather involved and will be given elsewhere. Preliminary calculations indicate that the results from \cite{TNL15} for $\epsilon_\mr{ave} \ll 1$ agree with those from the semiclassical analysis to $O(\epsilon_\mr{ave}^2)$.

We have also proposed and analyzed a novel interferometric scheme -- the SLIVER method of Sec.~3 -- for estimating the separation between the sources once their centroid has been located. The scheme alleviates Rayleigh's curse and provides superresolution at all values of source strengths. The essential ingredient of the scheme is an image inversion interferometer (III) -- a device that separates the symmetric and antisymmetric parts of the image-plane field relative to inversion about the centroid of the sources. The SLIVER method, unlike the binary SPADE and SPADE schemes, does not require coupling the image-plane field into one or more waveguides with mode profiles tailored to the PSF of the imaging system. As such, it appears to be more readily implementable in practice.

An explanation of the superlocalization effect of SLIVER for estimating $d \approx 0$ may be given in terms of photon statistics as follows. If both sources were superimposed at the centroid (so that $d=0$), the field $E_a(\brho)$ in the antisymmetric output port of the III  is identically zero despite their incoherence. If $d$ is small but nonzero, the mean photon number $\bar{N}_a$ is finite but still small as per Eq.~\eqref{Na}. The Fisher information on the mean of a Bose-Einstein distribution equals $[\bar{N}_a(\bar{N}_a +1)]^{-1}$ and hence is very large in this region of $d$. This extreme sensitivity to changes in $d$ is partially offset by the weak sensitivity of $\bar{N_a}$ on $d$ via $\gamma(d)$ which is close to zero for $d \approx 0$ (cf. the term in square brackets in Eq.~\eqref{septerms}). These two effects compensate each other to give a finite value of the Fisher information for separations close to zero. In the symmetric  port, on the other hand, the mean output is  large at $d \approx 0$ and the Fisher information is correspondingly small, so that the second effect  dominates and superlocalization is not obtained. Similar remarks apply to on-off detection in the two output ports. It is rather remarkable that appropriate linear-optics processing followed by spatially-unresolved on-off detection can alleviate the divergent MSE behavior plaguing ideal spatially-resolved image-plane photon counting for closely separated sources. Somewhat intriguingly, receivers using the same toolkit of operations (linear-optics processing with bulk on-off detection) also approach the quantum limits of binary \cite{Ken73,Dol73} and $M$-ary \cite{NGT14} communication with laser (coherent-state) light.
 
As mentioned in Sec.~3, various imaging modalities employing image inversion have been proposed and demonstrated previously in the microscopy literature \cite{SG06,WH07,WSH09,Wei+11,WBK+11,WEB+13,WBK+14,WBK+15}. The importance of the antisymmetric output of the III for improving lateral resolution has also been noticed, e.g., in \cite{WH07}. However, the claimed resolution enhancements in these works are in terms of the reduction of effective widths of the PSF by fixed factors, which in turn are related to the intensity patterns at the two outputs of the III (or their difference). Thus, from a statistical viewpoint, it appears that these works focus on the spatial structure of the first moment of the photocount random process (i.e, the intensity) at the two output channels of the interferometer and on the information about the separation $d$ extractable from it. In contrast, we have focused on spatially unresolved detection at the antisymmetric output port of the III and on the Cram\'er-Rao bound as the fundamental limit to the mean squared error of any (unbiased) estimate of the separation. By considering statistically optimal estimators, we can exploit the information available in the full probability distribution of the measurement rather than just its first moment. Such subtler estimators are necessitated by the limited number of photons that can be collected in most imaging applications. The maximum likelihood estimator, which approaches the Cram\'er-Rao bound in the limit of large $M$ \cite{VanTreesI}, was shown to yield performance in close agreement to the bound and sometimes exceeding it. 

In order to focus on the spatial aspects of the resolution problem, we have, as in \cite{TNL15}, suppressed the explicit temporal dependence of the field in our analysis. In effect, we have assumed that the light emitted by the sources is in a single quasimonochromatic temporal mode in each detection window. Such a situation can be easily realized using a pseudothermal light source \cite{Goo85Statistical}, but is in general unrealistic for many astronomical and biological imaging situations in which the coherence time of the emitted radiation is typically much shorter than the duration of the detection window. An approximate treatment of such cases can be given using the concept of the number of coherence cells \cite{Goo85Statistical}. This approach effectively redefines the number of shots $M$  according to the coherence properties of the light, with exact results obtainable in certain cases. These issues will be explored in detail elsewhere. 

 The effects of loss and non-unity quantum efficiency of the detectors can be incorporated in our analysis by appropriate scaling of the source strengths. Since spatially homogeneous losses simply scale the semiclassical field amplitudes, thermal states remain thermal states under such attenuation.Thus,  a system resolving sources of strength $\epsilon_\mr{ave}$ using a detector with quantum efficiency $\eta$ can be modeled as a system with source strength $\eta\epsilon_\mr{ave}$ and an ideal detector.  

We have assumed through most of this paper that the centroid of the sources has been located at the optical axis.  In optical astronomy applications, the centroid position may be available from previous observational records, or else a long observation time should be available to determine the centroid by, say, image-plane photon counting. In  microscopy applications, the time available for measurements may be more limited, e.g., because of photobleaching in fluorescent samples\cite{Kub13}, and lack of knowledge of the centroid position can be a significant impediment to implementing SLIVER.  To address this problem, one can first perform image-plane photon counting using a portion of the available light to determine the centroid position to within a small fraction of the PSF width before implementing SLIVER. The simulations in Sec.~5.3 indicate that the MSE of separation estimation should still be well below that achievable using image-plane photon counting. A more general study of the effects of centroid mismatch is best carried out in the framework of multi-parameter estimation theory and will be given elsewhere.
  
 Apart from exploring these issues, various extensions and refinements of the schemes of this paper suggest themselves. For example,  spatially-resolved detection in one or both output ports of the interferometer can only improve the performance of SLIVER. In any case, the results here should hopefully spur investigations into these issues in order to reap the superlocalization gain promised by the SLIVER method.

\section*{Acknowledgments}
This work is supported by the Singapore National Research Foundation under NRF Grant No.~NRF-NRFF2011-07.

\end{document}